\documentclass[aps,prb,amsmath,amssymb,twocolumn,floatfix]{revtex4-2}
\usepackage{graphicx}
\usepackage{dcolumn}
\usepackage{bm}
\usepackage{xcolor}
\usepackage{amsmath}
\usepackage{amssymb}   
\usepackage{epsfig}
\usepackage{epstopdf}
\usepackage{gensymb}

\begin{document}

\title{Gap structure of the non-symmorphic superconductor LaNiGa$_2$ probed by $\mu$SR}

\author{Shyam~Sundar,$^1$ M.~Yakovlev,$^2$ N.~Azari,$^2$ M.~Abedi,$^2$ D. M.~Broun,$^2$ H. U.~\"{O}zdemir,$^2$ S. R.~Dunsiger,$^{2,3}$ 
D.~Zackaria,$^4$ H.~Bowman,$^4$ P.~Klavins,$^4$ Y.~Shi,$^4$ V.~Taufour,$^4$  and J. E.~Sonier$^2$}

\affiliation{$^1$Scottish Universities Physics Alliance, School of Physics and Astronomy, University of St. Andrews, KY16 9SS, United Kingdom \\
$^2$Department of Physics, Simon Fraser University, Burnaby, British Columbia V5A 1S6, Canada \\
$^3$Centre for Molecular and Materials Science, TRIUMF, Vancouver, British Columbia V6T 2A3, Canada \\
$^4$Department of Physics and Astronomy, University of California, Davis, California 95616, USA}

\date{\today}
\begin{abstract}
We report muon spin rotation ($\mu$SR) measurements of the temperature dependence of the absolute value of the magnetic penetration depth and the
magnetic field dependence of the vortex core size in the mixed state of the non-symmorphic superconductor LaNiGa$_2$. 
The temperature dependence of the superfluid density is shown to be well described by a two-band model with strong interband coupling.
Consistent with a strong coupling of the superconducting condensates in two different bands, we show that the field dependence of the vortex core size resembles 
that of a single-band superconductor. Our results lend support to the proposal that LaNiGa$_2$ is a fully-gapped, 
internally antisymmetric nonunitary spin-triplet superconductor.  
\end{abstract}

\maketitle

\section{Introduction}

Superconductors exhibiting nontrivial topology have garnered significant attention in the field of condensed matter physics.
Of particular interest is the prediction that topological superconductors can host Majorana zero modes, which may
be used as building blocks for a quantum computer robust to environmental noise \cite{Sarma:2015,Sato:2017}.  
Recently, LaNiGa$_2$ has been identified as a potential topological superconductor via an investigation of newly synthesized high-quality single crystals \cite{Badger:2022}. 
The enhanced quality of the single crystals compared to previously studied polycrystalline samples has revealed a topological electronic band 
structure arising from non-symmorphic symmetries in the centrosymmetric space group of LaNiGa$_2$. In addition to the topological character, LaNiGa$_2$ is
believed to have an unconventional superconducting order parameter that breaks time-reversal symmetry (TRS). 
Zero-field (ZF) $\mu$SR measurements on polycrystalline LaNiGa$_2$ have detected the onset of weak spontaneous internal magnetic fields at the 
superconducting transition temperature ($T_c$), which is a signature of a TRS breaking superconductor \cite{Hillier:2012}. 
Furthermore, two nodeless superconducting gaps have been inferred from measurements of the temperature dependences of the specific heat, upper critical field and magnetic penetration depth 
(by a tunnel diode oscillator method) in polycrystalline LaNiGa$_2$ \cite{Weng:2016}. 
The occurrence of a nodeless two-gap superconducting state that breaks TRS has been explained by
an internally antisymmetric nonunitary triplet (INT) state in which there is both spin-up ($\uparrow \uparrow$) and spin-down ($\downarrow \downarrow$) pairing 
between electrons on two different atomic orbitals \cite{Weng:2016,Csire:2018,Ghosh:2020}.

The INT state has also been invoked to account for fully-gapped TRS breaking in the compositionally related noncentrosymmetric superconductor LaNiC$_2$. 
Evidence for a broken TRS superconducting state in LaNiC$_2$ has been observed by ZF-$\mu$SR measurements on a polycrystalline sample \cite{Hillier:2009},
although a ZF-$\mu$SR study of single crystals observed the occurrence of weak internal magnetic fields only below $T \! \sim \! 2/3T_c$ \cite{Sundar:2021}. 
Although different experiments have yielded inconsistent conclusions on the nature of the superconducting gap 
structure of LaNiC$_2$ \cite{Lee:1996,Pecharsky:1998,Iwamoto:1998,Chen:2013,Bonalde:2011,Katano:2017,Landaeta:2017},
the existence of two nodeless superconducting gaps was recently unambiguously verified by transverse-field (TF) $\mu$SR measurements of the low-field
magnetic penetration depth and low-temperature vortex core size in the mixed state of single crystals \cite{Sundar:2021}.
The two nodeless gaps manifest as a simultaneous crossover in the field dependences of an effective magnetic penetration depth and the vortex core size 
due to delocalization of the quasiparticle vortex-core states associated with the smaller gap.

In stark contrast to the high-quality single crystals of LaNiGa$_2$ in which topological superconductivity has recently been recognized \cite{Badger:2022}, 
the electronic specific heat of the polycrystalline LaNiGa$_2$ sample investigated in Ref.~\cite{Weng:2016} exhibits a rather broad superconducting transition
that is suggestive of significant inhomogeneity. The shape of the broad transition and its evolution with applied magnetic field is potentially an indication of
a double phase transition. This raises the possibility of nearly degenerate unconventional pairing states with critical temperatures that are split by non-magnetic disorder.
Since disorder can alter the temperature dependence of the physical quantities previously measured in polycrystalline LaNiGa$_2$, further evidence of there
being two nodeless superconducting gaps is needed.
Although a single sharp superconducting transition in the temperature dependence of the specific heat was reported in single crystalline samples \cite{Badger:2022}, 
no $\mu$SR studies have been done on LaNiGa$_2$ single crystals.

Here we report on a TF-$\mu$SR investigation of the superconducting energy-gap structure in LaNiGa$_2$ single crystals. 
From measurements that probe the magnetic field distribution in the vortex (mixed) state, we 
find that the temperature dependence of the normalized superfluid density for a magnetic field applied parallel to the $b$-axis is consistent 
with two nodeless superconducting energy gaps having magnitudes close to the sizes of the two apparent gaps in LaNiC$_2$. In contrast to LaNiC$_2$, however, 
the data for LaNiGa$_2$ indicates strong interband coupling. This is substantiated by the 
magnetic field dependence of the vortex core size, which as expected for strong coupling between two different band condensates, 
resembles that of a single isotropic gap superconductor in the clean limit. 

\section{Experimental Details}

Single crystals of LaNiGa$_2$ were grown using a Ga deficient self-flux technique, as described in Ref.~\cite{Badger:2022}. Magnetic susceptibility measurements
presented in the Supplemental Material \cite{SM} show that bulk superconductivity occurs at $T_c \! \sim \! 2$~K and all of the single crystals 
exhibit nearly full superconducting shielding fraction.  
The TF-$\mu$SR experiments were performed on the M15 surface muon beamline at TRIUMF, utilizing a top-loading dilution refrigerator. 
The sample consisted of multiple $b$-axis aligned LaNiGa$_2$ single crystals
arranged in a mosaic and mounted onto a pure Ag plate of dimensions 12.5~mm~$\times$~22~mm~$\times$~0.25~mm, as shown in the inset of  Fig.~\ref{fig1}(b).
The crystals covered $\sim \! 70$~\% of the Ag plate. To minimize the contribution to the TF-$\mu$SR signal from muons stopping outside the sample, three thin wafers of intrinsic GaAs were 
used to cover the exposed end of the Ag backing plate. We note that GaAs does not produce any detectable muon precession signal within the field range considered in our study.
The external magnetic field was applied parallel to the $b$ axis of the LaNiGa$_2$ single crystals 
and perpendicular to the initial muon-spin polarization ${\bf P}(t \! = \! 0)$. The magnetic field was applied above $T_c$ and the sample subsequently cooled down 
to the desired temperature in the superconducting state. The magnetic field distribution in the vortex state was probed for each temperature and magnetic field
by measuring the time evolution of the muon-spin polarization via detection of the decay positrons from an implanted ensemble of $\sim \! 15$ million positive muons.  
Further details on the TF-$\mu$SR method utilized in this study may be found in Ref.~\cite{Sonier:2000}.

\section{Data Analysis and Results}

\begin{figure}
\includegraphics[scale=0.45]{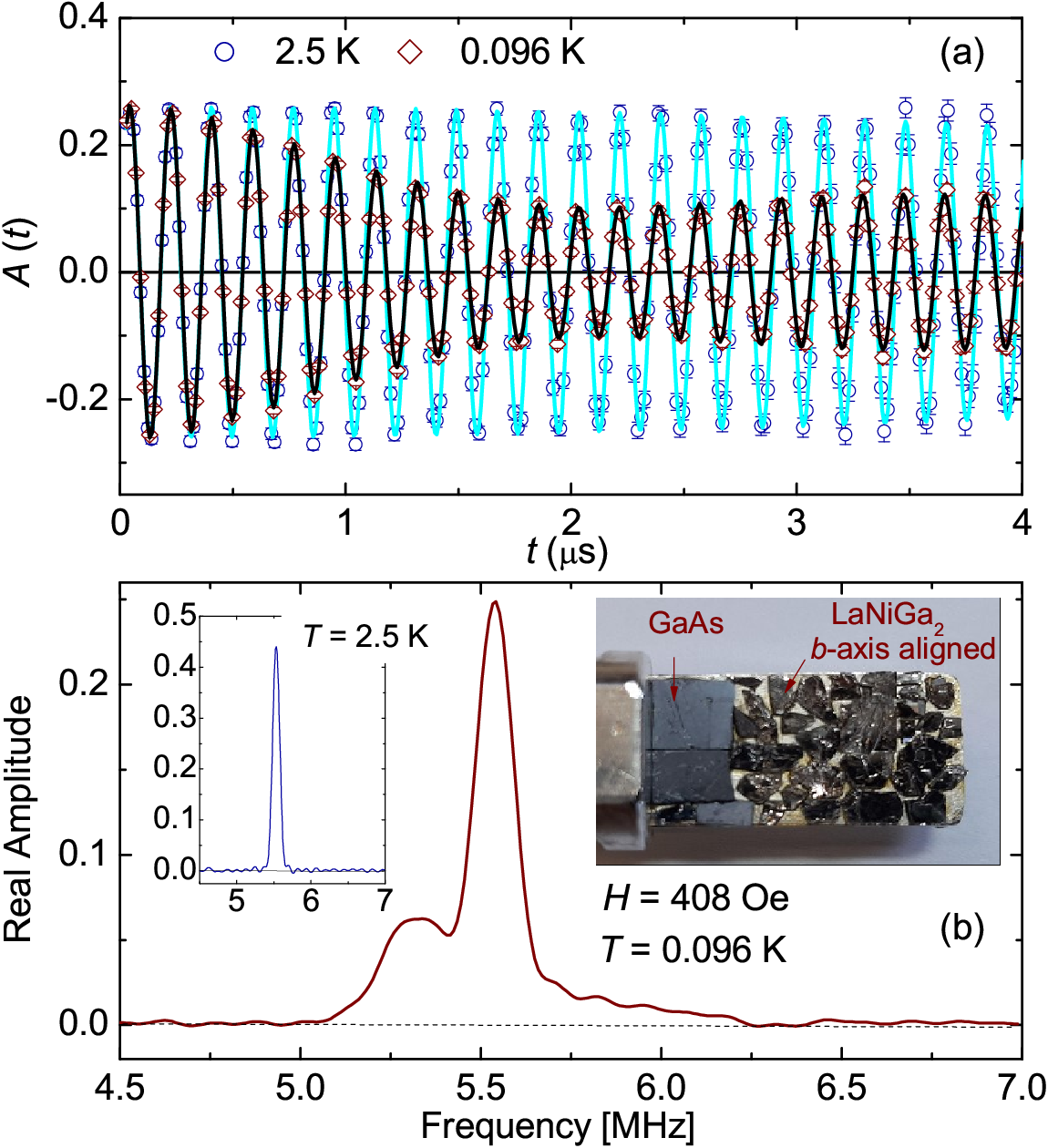}
\caption{(a) TF-$\mu$SR asymmetry spectra measured above and below $T_c$ for a magnetic field of $H \! = \! 408$~Oe. The oscillating curves through the data points represent fits to Eqs.~(\ref{eq:1}) and (\ref{eq:2}) for temperatures above and below $T_c$, respectively. (b) Fourier transform of the TF-$\mu$SR signal for $T \! = \! 0.096$~K. The large peak at 5.53~MHz is a result of muons stopping outside the sample. Left inset: Fourier transform of the TF-$\mu$SR signal for $T \! = \! 2.5$~K. Right inset: Photograph showing the LaNiGa$_2$ single crystals and GaAs wafers attached to the Ag backing plate, which is anchored to the Ag sample holder
of the dilution refrigerator. The single crystals are mounted with their $b$ axis aligned in the direction of the applied magnetic field, which is perpendicular to the plane of the Ag backing plate.}
\label{fig1}
\end{figure}
Figure~\ref{fig1}(a) displays representative TF-$\mu$SR asymmetry spectra recorded in the normal and superconducting states of LaNiGa$_2$ for an applied magnetic field of $H \! = \! 408$~Oe.
The weak damping of the TF-$\mu$SR asymmetry spectrum in the normal state at $T \! = \! 2.5$~K is due to the magnetic field distribution associated with nuclear dipole moments sensed by 
muons that stopped in the sample and the exposed area of the Ag backing plate. 
The larger depolarization rate of the TF-$\mu$SR signal at 0.096~K is a result of muons randomly sampling the spatial distribution of magnetic field 
generated by the vortex lattice below $T_c$. Figure~\ref{fig1}(b) shows Gaussian-apodized Fourier transforms of the TF-$\mu$SR signals. 
The Fourier transforms are only an approximate visual representation of the magnetic field distribution detected by the muons, because 
of the additional broadening
by the apodization used to remove the ringing and noise associated with the finite time range and reduced number of muon decay events at later times \cite{Sonier:2000}. 
Below $T_c$, the Fourier transform displays a distinct peak at the frequency corresponding to the applied field associated with muons that missed the sample and stopped in the Ag backing plate. This background peak is superimposed on an asymmetric lineshape 
that comes from muons that sensed the spatial distribution of field of the vortex lattice and the nuclear moments in the LaNiGa$_2$ single crystals.

Above $T_c$, the TF-$\mu$SR asymmetry spectrum is well described by the sum of two Gaussian-damped cosine functions
\begin{eqnarray} 
A(t)  & = & a_{\rm s} e^{-\sigma_{\rm s}^2 t^2} \cos(2 \pi \nu_{\rm s} t + \phi)  \nonumber \\
 & + &  a_{\rm bg}e^{-\sigma_{\rm bg}^2 t^2}\cos(2 \pi \nu_{\rm bg} t + \phi) \,.    
\label{eq:1}
\end{eqnarray}
The first term describes the signal from muons that stop in the LaNiGa$_2$ single crystals, while the second term accounts for the signal coming from muons stopping outside the sample.
The precession frequencies $\nu_i$ ($i \! =$ s, bg) are a measure of the corresponding mean local field $B_i \! = \! 2 \pi \nu_i/\gamma_\mu$ sensed by the muons, where $\gamma_\mu/2 \pi \! = \! 13.5539$~MHz/kG is the muon gyromagnetic ratio. The parameter $\phi$ is the initial phase of the muon spin polarization relative to the positron counters, which depends on the 
degree of Larmor precession of the muon spin in the applied field before reaching the sample. 

The TF-$\mu$SR signals below $T_c$ are well fit assuming the following modified analytical Ginzburg-Landau (GL) model \cite{Sonier:2004} for the
spatial variation of the internal magnetic field generated by the vortex lattice with supercurrents flowing in the $ij$~$(= \! ac)$ plane 
\begin{equation}
B({\bf r}) =  \sum_{ {\bf G}}
\frac{B_0 (1-b^4)e^{-i {\bf G} \cdot {\bf r}} \, \, u \, K_1(u)}{\lambda_{ij}^2 G^2 + 
\lambda_{ij}^4(n_{xxyy} \, G^4 + d \, G_x^2 G_y^2)} \, .
\label{eq:2}
\end{equation}
Here $b \! = \! B/B_{c2}$ is the reduced field (where $B_{c2} \! = \! \Phi_0/2 \pi \xi_{ij}^2$ is the upper critical magnetic field for a field applied perpendicular to the $ij$ plane), 
$B_0$ is the average internal magnetic field, 
{\bf G} are the reciprocal lattice vectors of the vortex lattice, $u^2 \! = \! 2 \xi_{ac}^2 G^2 (1 + b^4)[1-2b(1 - b)^2]$, $K_1(u)$ is a modified Bessel function, 
$n_{xxyy}$ and $d$ are dimensionless parameters arising from nonlocal corrections, and $\lambda_{ij}$ and $\xi_{ij}$ are the magnetic
penetration depth and GL coherence length. 
Equation~(\ref{eq:2}) accounts for potential changes in the vortex lattice geometry, ranging from hexagonal to square. 
The fits of the TF-$\mu$SR asymmetry spectra below $T_c$ were done by replacing the sample term in Eq.~(\ref{eq:1}) with
\begin{equation}
A_{\rm s}(t) = a_{\rm s}e^{-(\sigma_{\rm_n}^2 + \sigma_{\rm dis}^2) t^2} \sum_{ {\bf r}}\cos \left[ \gamma_\mu B({\bf r})  t + \phi \right]  \, ,
\label{eq:3}
\end{equation}
where $B({\bf r})$ is given by Eq.~(\ref{eq:2}).
The depolarization rate $\sigma_{\rm n}$ is the value of $\sigma_{\rm s}$ above $T_c$, which is due to the nuclear dipoles in the sample and is independent of temperature.
The sum is over real-space positions in an ideal periodic vortex lattice, while the depolarization rate $\sigma_{\rm dis}$ accounts for further broadening of the internal magnetic field 
distribution by frozen disorder in the vortex lattice. Based on previous $\mu$SR studies of type-II superconductors \cite{Sonier:2007}, we assumed $\sigma_{\rm dis}$ is 
proportional to $1/\lambda_{ac}^2$. 

Good fits to TF-$\mu$SR spectra recorded for an applied field of $H \! = \! 150$~Oe were achieved for all temperatures below $T_c$
assuming an hexagonal vortex lattice, with $n_{xxyy} \! = \! 0$ and $d \! = \! 0$. The same is true for TF-$\mu$SR spectra recorded for 
$H \! \leq \! 400$~Oe and $T \! = \! 0.096$~K. However, for $H \! = \! 800$~Oe and $T \! = \! 0.096$~K, a good fit could not be achieved with the 
assumption of an hexagonal vortex lattice. Instead, a good quality fit was achieved by lifting this constraint, yielding $n \! = \! (4.5 \! \pm \! 8.7) \! \times \! 10^{-8}$, 
$d \! = 0.29 \! \pm \! 0.18$, and $\beta \! = \!  90 \! \pm \! 17^{\circ}$, where $\beta$ is the acute angle of the rhombic unit cell of the vortex lattice. 
Even without assuming the vortex lattice structure, we could not achieve
a good fit of an additional TF-$\mu$SR asymmetry spectrum recorded for $H \! = \! 600$~Oe and $T \! = \! 0.096$~K. This is perhaps due to
a superposition of distorted hexagonal and square vortex lattices between the field-induced hexagonal-to-square lattice transition. In all likelihood
the vortex-lattice transition is driven by the anisotropy of the Fermi surface \cite{Takanaka:1971} in the $ac$ plane, which is close 
to having four-fold symmetry \cite{Badger:2022}. 

\begin{figure}[h]
\centering
\includegraphics[scale=0.9]{"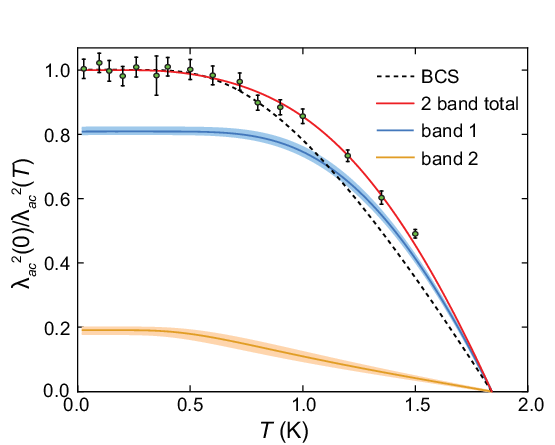"}
\caption{Temperature dependence of the normalized superfluid density, $\lambda^2_{ac}(0)/\lambda^2_{ac}(T)$. The circles denote the TF-$\mu$SR data points
for LaNiGa$_2$ in an applied magnetic field $H \! = \! 150$~Oe and the error bars give the standard error at each temperature. 
The dashed curve is the normalized superfluid density from single-band BCS theory \cite{BCS:1957} assuming the reduced value of
 $T_c = 1.84$~K for $H \! = \! 150$~Oe determined from bulk magnetic susceptibility measurements presented in the Supplemental Material \cite{SM}. 
 The upper solid curve is the total normalized superfluid density in the two-band model for the density of states parameter $n_1 = 0.2$, with contributions
 from the individual bands shown below it.  Shaded areas denote the 1-$\sigma$ uncertainty regions associated with the model fit. The fit parameters are given in the Appendix.}
\label{fig2}
\end{figure}

As one of few techniques able to reliably probe the \emph{absolute value} of the magnetic penetration depth, $\mu$SR correctly obtains the shape of the superfluid density
$\rho_s(T)$.  We can then infer information about the magnitude, anisotropy and temperature dependence of the energy gap, and 
the strength and nature of the pairing interaction, by fitting generalized Bardeen-Cooper-Schrieffer (BCS) models to the superfluid density.
For LaNiGa$_2$, the normalized low-field superfluid density, $\rho_s(T)/\rho_s(0) \! = \! \lambda^2_{ac}(0)/\lambda^2_{ac}(T)$, 
where $\lambda_{ac}(0) \! = \! 1516 \! \pm \! 4$~\AA, is shown in Fig.~\ref{fig2}.
We can immediately see from Fig.~\ref{fig2} that the superfluid density in LaNiGa$_2$ is distinctly different from that of a single-band BCS superconductor, in particular having larger magnitude 
in the upper half of the superconducting temperature range.  We will later see that this is the experimental signature of strong interband pairing
in the presence of a density of states imbalance between the bands.
The value of $\lambda_{ac}(0)$ is much smaller than the $\mu$SR-determined value of the effective magnetic penetration depth 
$\lambda(0) \! = \! 3500 \! \pm \! 100$~\AA~ in a polycrystalline LaNiGa$_2$ sample \cite{Hillier:2012}.
For polycrystalline samples, the internal magnetic field distribution measured by $\mu$SR in the vortex state is an average over all orientations of the external magnetic field
with respect to the crystalline axes. Consequently, $\lambda(0)$ is an average of the $a$-axis, $b$-axis and $c$-axis magnetic penetration depths, which have been estimated
to be $\lambda_a \! = \! 1740$~\AA, $\lambda_b \! = \! 5090$~\AA~ and $\lambda_c \! = \! 1890$~\AA~ from a combination of thermodynamic critical field $H_c(T)$ 
and anisotropic upper critical field $H_{c2}(T)$ data for LaNiGa$_2$ single crystals \cite{Badger:2022}. 
We note that the value of $\lambda_{ac}(0)$ determined here by fits of the TF-$\mu$SR asymmetry spectra with Eq.~(\ref{eq:2}) agrees with the
value $\lambda_{ac} \! = \! 1512 \! \pm \! 6$~\AA~ calculated from the second moment $\langle \Delta B^2 \rangle$ of the internal magnetic field
distribution probed by the muons, where $\langle \Delta B^2 \rangle$ is determined from fits of the TF-$\mu$SR spectra below $T_c$ to the sum of 
four Gaussian-damped cosine functions (see Fig.~S2 in the Supplemental Material \cite{SM}). This alternative analysis method \cite{Brandt:88} has been employed 
extensively to determine the absolute value of the magnetic penetration depth from $\mu$SR data on type-II superconductors, but does not yield information on
the vortex-core size. 

To capture the behavior of the superfluid density, we turn to a two-band BCS model, which can be viewed as the low energy effective theory of a superconductor displaying significant anisotropy of pairing over the Fermi surface, but without nodal lines or point nodes in the energy gap. The two-band model we use has been presented 
in detail in Ref.~\cite{Kogan:2009} and has been successfully used to describe the superconductivity in the non-centrosymmetric material LaNiC$_2$ 
(Ref.~\cite{Sundar:2021}) --- although we note from the outset that the inferences drawn about the pairing interactions in LaNiGa$_2$ are quite different from 
those in LaNiC$_2$, likely reflecting the different symmetry-allowed superconducting order parameters in the two materials.

In a nutshell, the two-band model partitions the Fermi surface into two disparate pieces, and allows for pairing interactions within the bands (intraband pairing, characterized by interaction parameters 
$\lambda_{11}$ and $\lambda_{22}$) and pairing interactions between the bands (interband pairing, characterized by interaction parameters $\lambda_{12} \! = \! \lambda_{21}$).   As shown in
Refs.~\cite{Kogan:2009} and \cite{Sundar:2021}, the usual situation is for the intraband pairing in one of the bands to dominate, and therefore 
for the interband interaction $\lambda_{12}$ (or $\lambda_{21}$) to be significantly smaller than the dominant intraband interaction.  
However, when we solve the gap equation for this model and fit to the superfluid density of LaNiGa$_2$, as described in the
Appendix, we find that the interband  interaction dominates in LaNiGa$_2$ ($\lambda_{12} \! \gg \! \lambda_{11}, \lambda_{22}$) --- something that has in fact been 
proposed theoretically for the unusual type of superconductivity thought to exist in LaNiGa$_2$ \cite{Weng:2016,Ghosh:2020}. 

As we can see in Fig.~\ref{fig2}, the two-band model provides an excellent fit to the normalized superfluid density, $\lambda^2_{ac}(0)/\lambda^2_{ac}(T)$, and is readily
able to capture an unusual feature of the data: the significant enhancement of superfluid density in the upper half of the temperature range compared to the single-band BCS curve. 
We note, and it can be seen in Ref.~\cite{Kogan:2009}, that the usual situation in the two-band model is for the total superfluid density to fall below the single-band BCS limit.  
However, the enhancement of the superfluid density observed in LaNiGa$_2$ is a natural consequence of predominantly interband pairing, combined with an imbalance in the density of states of the two bands.
Since the superfluid density in each band is purely a function of $\Delta/T$, it is possible for the total superfluid density to exceed the BCS limit, as long as one of the gaps is enhanced above the BCS value 
and the relative superfluid weight ($\gamma$ parameter in our model) is balanced in favour of the band with the larger gap.  
The energy gaps obtained from fitting the two-band model are plotted in the Appendix (see Fig.~\ref{fig:gaps}) and indeed reveal such an enhancement of the dominant gap. 

BCS-type superconductivity is characterized by
thermally activated behaviour in low temperature properties such as the superfluid density: accordingly, the leading low temperature behaviour of the superfluid density ({\it i.e.}, the temperature range over 
which significant $T$ dependence develops) is set by the  magnitude of the subdominant gap.  This is particularly apparent in the lower, band-2 curve in the 
decomposition of the normalized superfluid density in Fig.~\ref{fig2}. 
From this we obtain a gap ratio of  $\Delta_2(0)/k_B T_c = 1.25$ for the subdominant band. In the two-band model, the magnitude of the dominant gap is usually set by $T_c$, but 
here we find that the dominant gap is larger, with $\Delta_1(0)/k_B T_c = 2.7$, much bigger than the BCS gap ratio of 1.76. 
We note that the gap ratios we find here are somewhat comparable to the values 1.29 and 2.04 deduced from superfluid density measurements of polycrystalline LaNiGa$_2$ in Ref.~\cite{Weng:2016}.
Within the two-band model, the enhancement above the usual BCS gap ratio is a direct consequence of the strong interband interaction. Finally, the fits reveal that the intraband pairing in band 1 
is intrinsically repulsive --- {\it i.e.}, it is only superconducting because of the interband interaction. These results are summarized in Table~1 in the Appendix.

\begin{figure}
\includegraphics[scale=0.48]{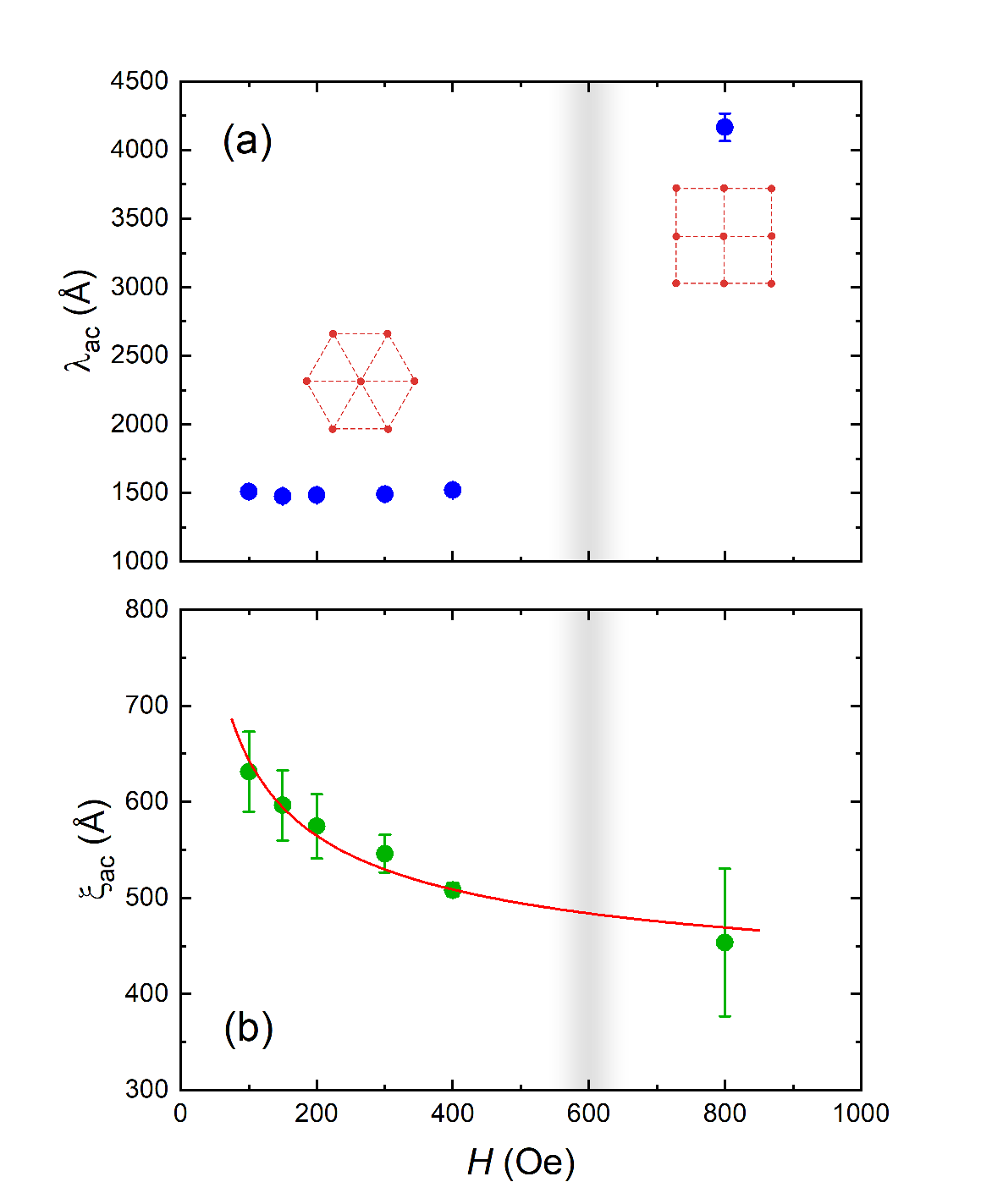}
\caption{(a) Magnetic field dependence of $\lambda_{ac}$ in LaNiGa$_2$ for $T \! = \! 0.096$~K from fits indicating
hexagonal and square vortex lattices for $H \! \le \! 400$~Oe and $H \! = \! 800$~Oe, respectively. 
(b) Magnetic field dependence of $\xi_{ac}$ in LaNiGa$_2$. The solid curve through the data points is a fit to $\xi_{ac} \! = \! a \! + \! b/\sqrt{H}$,
where $a \! = \! 374 \! \pm \! 19$~\AA~ and $b \! = \! 2704 \! \pm \! 269$~\AA$\cdot$Oe$^{1/2}$.}
\label{fig3}
\end{figure}
\begin{figure}
\includegraphics[scale=0.48]{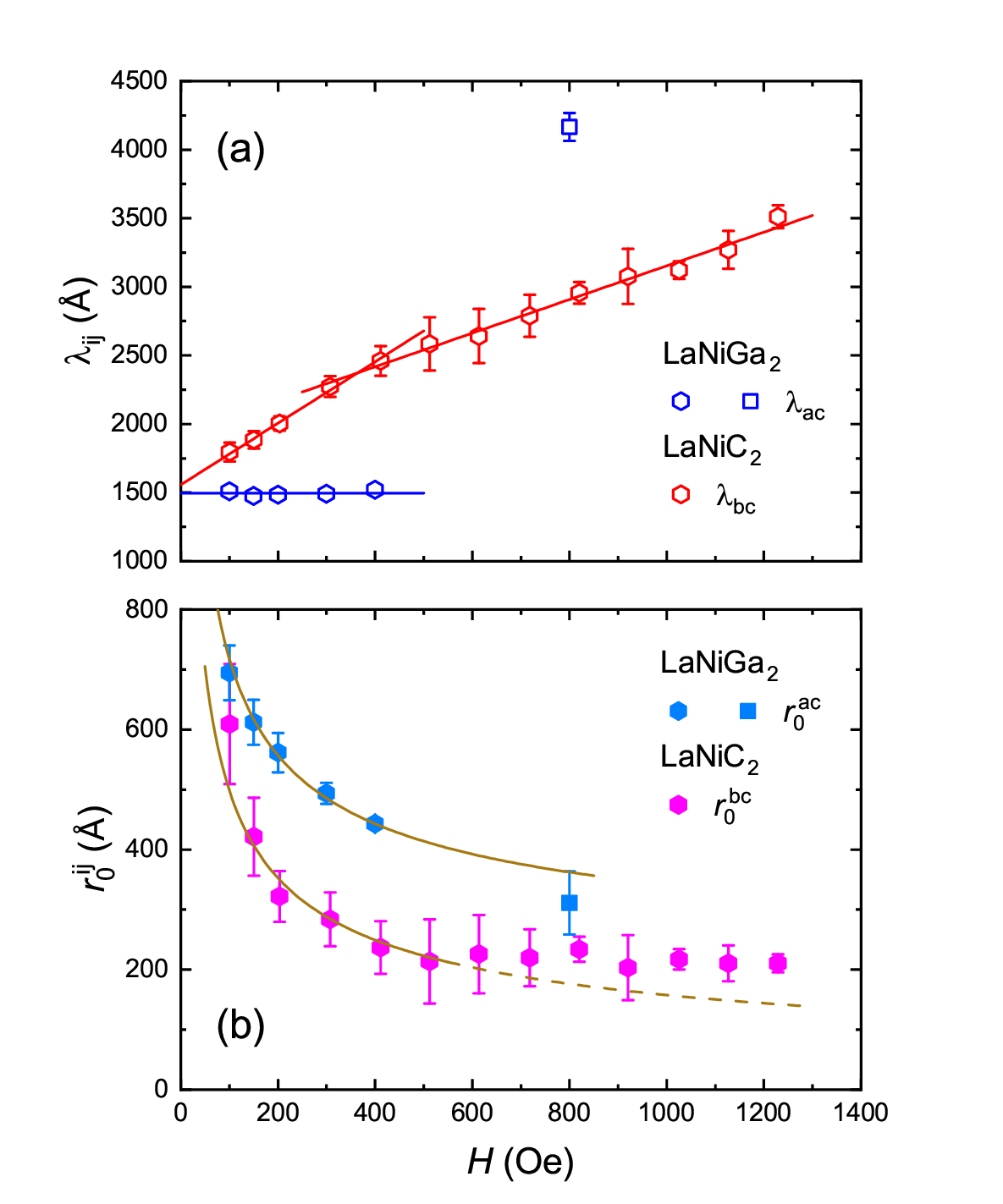}
\caption{Comparison of the magnetic field dependences of the fitted values of (a) $\lambda_{ac}$ and (b) $\xi_{ac}$ in LaNiGa$_2$ for $T$ = 0.096 K and 
${\bf H} \! \parallel \! {\bf b}$ with the values of $\lambda_{bc}$ and $\xi_{bc}$ in LaNiC$_2$ for $T$ = 0.05 K and ${\bf H} \! \parallel \! {\bf a}$ 
from Ref.~\cite{Sundar:2021}. The solid lines through the data point in (a) are guides to the eye. The curves through the data points in (b) are
fits to $r^{ij}_0 \! = \! a \! + \! b/\sqrt{H}$, where $a \! = \! 168(28)$~\AA~ and $b \! = \! 5504(452)$~\AA$\cdot$Oe$^{1/2}$ for the core size
$r^{ac}_0$ in LaNiGa$_2$. The fit to the field dependence of the core size $r^{bc}_0$ for LaNiC$_2$ is to the data at $H \! \leq \! 400$~Oe,
where $a \! = \! 0$~\AA~ and $b \! = \! 4986(228)$~\AA$\cdot$Oe$^{1/2}$.  
The hexagonal and square data symbols for LaNiGa$_2$ indicate the geometrical arrangement of the vortices.}
\label{fig4}
\end{figure}

Figure~\ref{fig3} shows the dependence of $\lambda_{ac}$ and $\xi_{ac}$ on the applied magnetic field for $T \! = \! 0.096$~K. Below $H \! = \! 400$~Oe, which corresponds
to the reduced field $b \! \sim \! 0.4$, $\lambda_{ac}$ is independent of the applied field. This is in stark contrast to the rapid increase in 
$\lambda_{ij}$ ($ij \! = \! ab$, $bc$ or $ac$) with increasing field at low $b$ determined by $\mu$SR in superconductors 
with gap nodes \cite{Sonier:2007,Kadono:2004} or strong gap anisotropy \cite{Ohishi:2002,Price:2002}.
A strong $H$ dependence of $\lambda_{ij}$ determined by $\mu$SR may also occur at low $b$ in weakly-coupled 
two-band superconductors, due to a faster suppression of the contribution from the weaker small-gap band to the superfluid density \cite{Callaghan:2005,Williams:2009,Weyeneth:2010}. 

In general, an increase in the $\mu$SR-determined value of $\lambda_{ij}$ with increasing magnetic field does not necessarily imply a field-induced 
change in the superfluid density, but may instead reflect a failure of the assumed model for $B({\bf r})$ to adequately describe changes in the decay of 
magnetic field around the vortices as the overlap between vortices increases \cite{Sonier:2007}.
We also note that not all $\mu$SR studies assume a theoretical model for $B({\bf r})$. A strong decrease in the square root of the second moment of the 
local magnetic field distribution measured by $\mu$SR in the vortex state of various superconductors
has been interpreted in terms of a field-induced reduction of the superfluid density or increase in the ``true'' magnetic penetration depth
\cite{Serventi:2004,Khasanov:2007,Khasanov:2008,Luetkens:2008}.  
Nevertheless, the $H$-independent behavior of $\lambda_{ac}$ in Fig.~\ref{fig3}(a) up to $H \! = \! 400$~Oe indicates that the field does
not induce a change in the superfluid density below $b \! \sim \! 0.4$.
 
The absence of any change in $\lambda_{ac}$ for LaNiGa$_2$ below $b \! \sim \! 0.4$ rules out the presence of gaps nodes and is in stark contrast to the strong field dependence 
of $\lambda_{bc}$ in the noncentrosymmetric superconductor LaNiC$_2$ [see  Fig.~\ref{fig4}(a)]. 
In LaNiC$_2$, $\lambda_{bc}$ exhibits an $H$-linear dependence with a slope change above $b \! \sim \! 0.2$, which is indicative of two superconducting gaps.
By contrast, the field dependence of $\lambda_{ac}$ in LaNiGa$_2$ closely resembles the behavior of the effective magnetic penetration depth measured by $\mu$SR
in V$_3$Si, which also exhibits an hexagonal-to-square vortex lattice transition accompanied by a change in the fitted value of $\lambda_{ij}$ \cite{Sonier:2004}.
Recent experiments suggest V$_3$Si has two distinct nodeless superconducting gaps \cite{Cho:2022,Ding:2023}.
Consequently, the constant value of $\lambda_{ac}$ below $b \! \sim \! 0.4$ for LaNiGa$_2$ does not rule out the occurrence of two gaps. 

Less ambiguous is the meaning of the field dependence of $\xi_{ac}$ shown in Fig.~\ref{fig3}(b), which except at low $H$, 
more or less tracks the dependence of the vortex core size on magnetic field \cite{Sonier:2007}.
The vortex core size ($r_0$) is accurately determined by calculating the absolute value of the supercurrent density profile $\lvert j({\bf r}) \rvert$ from the
experimental $B({\bf r})$ and defining $r_0$ to be the distance from the core center along the nearest-neighbor vortex direction to the
peak in $\lvert j({\bf r}) \rvert$ \cite{Sonier:1997}. It has been shown that the field dependence of $r_0$ measured by $\mu$SR
precisely accounts for the field dependence of the electronic thermal conductivity measured in V$_3$Si, 2H-NbSe$_2$ and LuNi$_2$B$_2$C \cite{Sonier:2007}.
The physical picture is that the vortex core size shrinks with increasing magnetic field as a result of an increased intervortex transfer of quasiparticles
at higher fields, where the distance between vortices is smaller \cite{Ichioka:1999a,Ichioka:1999b}. 
The increased delocalization of the bound quasiparticle vortex core states is detected in the electronic thermal conductivity measurements \cite{Boaknin:2003}.
As can be seen in Figs~\ref{fig3}(b) and \ref{fig4}(b), the field dependences of $\xi_{ac}$ and the vortex core size $r^{ac}_0$ for LaNiGa$_2$ are approximately
described by an expression of the form $a \! + \! b/\sqrt{H}$, which is the predicted behavior of the
vortex core size in clean isotropic-gapped BCS superconductors \cite{Kogan:2005}. This is distinct from the two-band behavior of the vortex core size in
LaNiC$_2$ [see Fig.~\ref{fig4}(b)] and that reported earlier for 2H-NbSe$_2$ \cite{Callaghan:2005}, where the core size rapidly decreases
with increasing field and becomes independent of $H$ at higher fields ---  although admittedly there is insufficient data for LaNiGa$_2$ at higher field (because of
the field-induced vortex-lattice transition) to completely rule out a crossover to $H$-independent behavior.  

The gap values determined from the analysis of the data in Fig.~\ref{fig2} ($\Delta_1 \! = \! 0.43$~meV and $\Delta_2 \! = \! 0.20$~meV) are comparable to
the size of the two gaps in LaNiC$_2$ ($\Delta_1 \! = \! 0.42$~meV and $\Delta_1 \! = \! 0.18$~meV) determined in Ref.~\cite{Sundar:2021}.
Hence, the different field dependences of $\lambda_{ij}$ and $r^{ij}_0$ exhibited by LaNiGa$_2$ and LaNiC$_2$ in Figs.~\ref{fig4}(a) and \ref{fig4}(b) are not due to dissimilar sizes of the two gaps. 
Interestingly, the low-temperature $H \! \rightarrow \! 0$ extrapolated value 
of $\lambda_{ac}$ for LaNiGa$_2$ is close to the low-temperature $H \! \rightarrow \! 0$ extrapolated value of $\lambda_{bc}$ for LaNiC$_2$, and the low-temperature, 
low-field values of $r^{ac}_0$ for LaNiGa$_2$ and $r^{bc}_0$ for LaNiC$_2$ are also comparable. It is possible that with increasing field a significant anisotropy develops
for the ratios $\lambda_{ac}/\lambda_{bc}$ and $r^{ac}_0/r^{bc}_0$, due to two-gap or anisotropic single-gap superconductivity. However, the $H$-independent behavior 
exhibited by $\lambda_{ac}$ in LaNiGa$_2$ for $H \! \leq \! 400$~Oe suggests that this is unlikely to be the origin of the different field dependences of $\lambda_{ij}$ and
$r^{ij}_0$ for the two compounds displayed in Figs.~\ref{fig4}(a) and \ref{fig4}(b).

Theoretically, it has been shown that in two-band $s$-wave superconductors with a significant difference in the magnitudes of the two gaps, the spatial variation of the superconducting 
order parameter near the vortex core is the same in both bands when there is strong interband coupling \cite{Silaev:2011,Ichioka:2017,Vargunin:2019}. Consequently, the 
magnetic field dependence of the vortex-core size in the different bands is the same and resembles that of a single-band superconductor.
Thus, the single-gap like field dependences of $\lambda_{ac}$ and $r^{ac}_0$ for LaNiGa$_2$ may be attributed to the strong interband coupling
deduced from the analysis of the temperature dependence of the normalized superfluid density.

To summarize, we have used $\mu$SR to investigate the temperature dependence of the superfluid density and magnetic field dependence of the vortex core
size in LaNiGa$_2$ single crystals. Together they are explained by two-band nodeless gap superconductivity with strong interband coupling. 
This lends support to the applicability of the proposed INT pairing state \cite{Weng:2016,Csire:2018,Ghosh:2020} to superconductivity in LaNiGa$_2$, which attributes the occurrence 
of the two gaps to the two spin-triplet states, $\uparrow \uparrow$ and $\downarrow \downarrow$, associated with Cooper pairing of electrons on two different Ni orbitals. 
The strong interband pairing inferred from our experimental results is notably inherent in the INT state. 


\section{Appendix: Two-Band Superconductivity}

\subsection{Two-band BCS theory}

In the Matsubara formalism, the temperature dependent gap equation for a weak-coupling superconductor is

\begin{equation}
\Delta_{\mathbf{k}}=2 \pi T N_0 \sum_{\omega_{n}>0}^{\omega_{0}}\left\langle V_{\mathbf{k}, \mathbf{k}^{\prime}} \frac{\Delta_{\mathbf{k}^{\prime}}}{\sqrt{\Delta_{\mathbf{k}^\prime}^{2} + \hbar^{2} \omega_{n}^{2}}}\right\rangle_{\!\!\mathrm{FS}}
\end{equation}
where $\omega_n = 2 \pi T (n + \frac{1}{2})$ are the fermionic Matsubara frequencies, $\Delta_\mathbf{k}$ is the gap parameter at wave vector $\mathbf{k}$, $N_0$ is the two-spin density of states, $V_{\mathbf{k}, \mathbf{k}^{\prime}}$ is the pairing interaction, $\langle...\rangle_\mathrm{FS}$ denotes an average over the Fermi surface and $\omega_0$ is a high frequency cutoff.  

\begin{figure}[h]
    \centering
    \includegraphics[width = 3 in]{"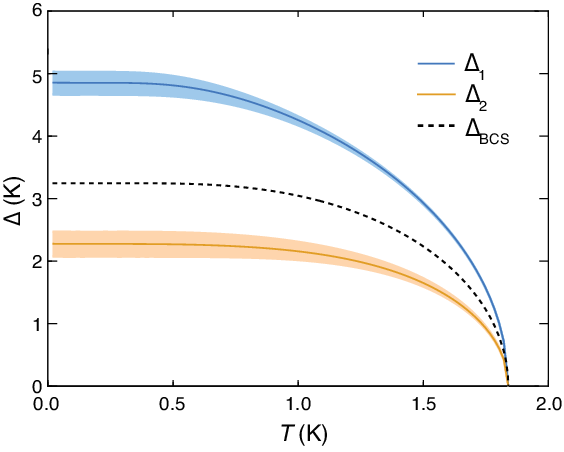"}
    \caption{Temperature dependence of the energy gaps, $\Delta_1$ and $\Delta_2$, for $T_c \! = \! 1.84$~K in the two-band model, for density of states parameter $n_1 = 0.2$. 
    Shaded areas denote the 1-$\sigma$ uncertainty regions associated with the model fit.  The zero-temperature gap ratios, $\Delta(0)_i/k_B T_c$, are 2.7 and 1.25, respectively.}
    \label{fig:gaps}
\end{figure}

The two-band superconductor describes situations in which the gap variation over the Fermi surface is approximately bimodal and can be approximated by two distinct gap scales, $\Delta_1$ and $\Delta_2$, one for each band.  As discussed in Ref.~\cite{Kogan:2009}, the Fermi surface average is replaced by a sum over bands, and the pairing interaction is parameterized by a $2 \times 2$ symmetric matrix $\lambda_{\mu \nu}$, with the diagonal terms $\lambda_{11}$ and $\lambda_{22}$ describing intraband pairing, and the off-diagonal terms $\lambda_{12} = \lambda_{21}$ the interband interaction.  The gap equation then takes the simplified form
\begin{equation}
\Delta_{\nu}=\sum_{\mu=1,2} n_{\mu} \lambda_{\nu \mu} 2 \pi T \sum_{\omega_{n}>0}^{\omega_0}\frac{ \Delta_{\mu}}{\sqrt{\Delta_{\mu}^{2} +\hbar^{2} \omega_{n}^{2}}}\;,
\label{eqn:gapTemperature}
\end{equation}
where the relative densities of states for each band, $n_\mu$, obey $n_1 + n_2 = 1$.
For a given choice of parameters $\{n_1,\lambda_{11},\lambda_{22},\lambda_{12}\}$, Eq.~(\ref{eqn:gapTemperature}) is solved numerically, from which we obtain the temperature dependence of the gap parameters $\Delta_1$ and $\Delta_2$, as shown, for example, in Fig.~\ref{fig:gaps}.

\subsection{Superfluid Density}

Superfluid density is a thermal equilibrium property of the superconductor and is readily obtained within the Matsubara formalism once the energy gaps are known.  For band $\nu$, the normalized superfluid density is 
\begin{equation}
\rho_{\nu}(T)= \frac{\lambda_\nu^2(0)}{\lambda_\nu^2(T)} = \sum_{\omega_{n}>0}\frac{\Delta_{\nu}^{2} }{\left(\Delta_{\nu}^{2} +\hbar^2\omega_n^{2}\right)^{3/2}}\;.
\label{eqn:SFDensity}
\end{equation}
The total normalized superfluid density is a weighted sum of the contributions from each band,
\begin{equation}
\rho(T)=\gamma \rho_{1}(T)+(1-\gamma) \rho_{2}(t)\;,
\label{eqn:totalSF}
\end{equation}
where the weighting factor $0 < \gamma < 1$ is determined by the plasma frequency imbalance balance the bands. Note that $\gamma$ is distinct from the density of states parameter $n_1$, as it includes Fermi velocity information: 
\begin{equation}
\gamma=\frac{n_{1} v_{1}^{2}}{n_{1} v_{1}^{2}+n_{2} v_{2}^{2}}\;,
\end{equation}
where $v_1$ and $v_2$ are the rms Fermi velocities of the two bands.

\subsection{Fitting procedure and results}

A least-squares optimization is used to search for best-fit parameters in the four-dimensional parameter space $\{n_1,\lambda_{11},\lambda_{22},\lambda_{12} \}$.  For each parameter choice, the band-specific energy gaps and superfluid densities are determined at each of the experimental temperatures via numerical solution of Eqs.~(\ref{eqn:gapTemperature}) and (\ref{eqn:SFDensity}). As shown in Eq.~(\ref{eqn:totalSF}), the total superfluid density is a weighted combination of the band-specific superfluid densities.  While the weighting coefficient $\gamma$ is formally an additional fit parameter, a closed-form expression exists for its optimal value, so that it need not be included in the minimization search.  $\gamma_\mathrm{opt}$ is found by minimizing the $\chi^2$ merit function
\begin{equation}
\begin{aligned}
\chi^{2} &=\left|\frac{\vec{\rho}_{\text {expt }}-\vec{\rho}_{\text {model }}}{\vec{\sigma}}\right|^{2} \\
&=\left|\frac{\vec{\rho}_{\text {expt }}-\vec{\rho}_{2}-\gamma \Delta \vec{\rho}}{\vec{\sigma}}\right|^{2} \\
&=\frac{\gamma^{2}|\Delta \vec{\rho}|^{2}-2 \gamma \Delta \vec{\rho} \cdot\left(\vec{\rho}_{\text {expt }}-\vec{\rho}_{2}\right)+\left|\vec{\rho}_{\text {expt }}-\vec{\rho}_{2}\right|^{2}}{|\vec{\sigma}|^2}
\end{aligned}
\end{equation}
where $\Delta \vec{\rho}=\vec{\rho}_{1}-\vec{\rho}_{2}$. Here the vector quantities encode the discrete temperature dependences of the various quantities, including experimental and model superfluid densities, and the measurement errors $\vec \sigma$. Minimizing with respect to $\gamma$ we obtain
\begin{equation}
\gamma_{\mathrm{opt}}=\frac{\Delta \vec{\rho} \cdot\left(\vec{\rho}_{\mathrm{expt}}-\vec{\rho}_{2}\right)}{|\Delta \vec{\rho}|^{2}}.
\end{equation}

\begin{table}[h!]
  \begin{center}
    \label{tab:table1}
    \begin{tabular}{|c||c|c||c|c|} 
    \hline
      fit parameter & $n_1 = 0.1$ & uncertainty &$n_1 = 0.2$ & uncertainty  \\
      \hline
      $\lambda_{11}$ & -0.15 & $\pm 0.039$ & -0.14 & $\pm 0.005$ \\
      $\lambda_{22}$ & 0.35 & $\pm 0.003$ & 0.27 & $\pm 0.026$ \\
      $\lambda_{12}$ & 1.54 & $\pm 0.26$ & 1.63 & $\pm 0.34$ \\
      $\gamma_\mathrm{opt}$ & 0.71 & & 0.81 &   \\ \hline
    \end{tabular}
    \caption{Best-fit parameters and their uncertainties, for $n_1 = 0.1$ and $n_1 = 0.2$.}
  \end{center}
\end{table}

In practice, the optimization depends only weakly on the choice of density of states parameter $n_1$, which we set to fixed values. 
We present results in Table~1 for $n_1 \! = \! 0.1$ and $n_1 \! = \! 0.2$, noting that fits are noticeably worse for $n_1 \! < \! 0.1$ and $n_1 \! > \! 0.2$.   
Figures \ref{fig2} and \ref{fig:gaps} show the fits and gaps for $n_1 \! = \! 0.2$, which are practically indistinguishable from the fits for $n_1 = 0.1$.  Note that while the $\lambda_{11}$ parameter appears to vary strongly between the two cases, it is the combination $n_1 \lambda_{11}$ that determines the intraband pairing strength in the first band, and this combination remains approximately constant.  From this we conclude that while the intrinsic pairing strength in the dominant band is actually replusive in LaNiGa$_2$, it is overwhelmed 
by the very significant interband contribution to its pairing. We reiterate that the experimental signature of this is the enhancement of $\rho_s(T)$ in the upper half of the 
superconducting temperature range, well in excess of the single-band BCS behaviour. We note that this is quite different from LaNiC$_2$, which shows no such 
enhancement, and has far less significant interband pairing.
 
\acknowledgments{We thank the personnel of the Centre for Molecular and Materials Science at TRIUMF for technical assistance
with our $\mu$SR measurements. J.E.S. and S.R.D. acknowledge support from the Natural Sciences and Engineering Research Council of Canada (PIN: 146772).
S.S. acknowledges support from the Engineering and Physical Sciences Research Council (EPSRC) through grant EP/P024564/1.
The sample synthesis at UC Davis was supported by the UC Laboratory Fees Research Program (LFR-20-653926). H.B. was supported by the NSF-REU program PHY-2150515.}

\bibliographystyle{apsrev}

\newpage

\setcounter{table}{0}
\setcounter{figure}{0}
\renewcommand{\thefigure}{S\arabic{figure}}
\hspace{12mm}
{\bf SUPPLEMENTAL MATERIAL}\\
\vspace{\baselineskip}

\noindent
{\bf Sample Characterization}\\

To characterize the LaNiGa$_2$ single crystals used in the mosaic for the $\mu$SR experiment, we estimated the superconducting shielding fraction 
from zero-field cooled (ZFC) bulk magnetic susceptibility measurements. The data is reported in Table~S1. All of the single crystals show nearly full 
magnetic shielding. Given the irregular shape of the single crystals, we did not correct for demagnetization effects. 
However, the measurements were performed with the applied magnetic field along the basal plane of each single crystal, where demagnetization effects are minimal.
The superconducting transition temperature ($T_c$) was measured for a few of the single crystals. As shown in Fig.~S1(a) the onset of the transition in bulk magnetic susceptibility is above 2~K, and nearly complete below 1.9~K. The relatively large magnetic shielding observed in the field cooled (FC) measurements indicate relatively low vortex pinning, which is an indication of good sample quality.

Figure~S1(b) shows bulk magnetic susceptibility measurements performed on one of the single crystals for different magnetic fields applied parallel to the $b$ axis.
From these measurements the value of $T_c$ for a magnetic field of $H \! = \! 150$~Oe is estimated to be $T \! = \! 1.84$~K, as shown in Fig.~S1(c). This value
is in agreement with the resistivity measurements on LaNiGa$_2$ single crystals reported in Ref.~\cite{Badger:2022}.\\

\begin{table}
\caption{Superconducting shielding fraction of the LaNiGa$_2$ single crystals for $T \! = \! 1.85$~K and $H \! = \! 10$~Oe.}
\begin{ruledtabular}
\begin{tabular}{cc}
 {\bf Single Crystal} & {\bf Superconducting shielding fraction} \\ \hline
 1 & 95~\% \\
 2 & 98.6~\% \\
 3 & 89~\% \\
 4 & 88~\% \\
 5 & 88~\% \\
 6 & 89~\% \\
 7 & 89~\% \\
 8 & 87.8~\% \\
 9 & 88.5~\% \\
10 & 94~\% \\
11  & 98~\% \\
12 & 99~\% \\
13 & 92~\% \\
14 & 91~\% \\
15 & 100~\% \\
16 & 92~\% \\
17 & 97~\% \\
18 & 91~\% \\
19 & 100~\% \\
20& 96~\% \\
21 & 95~\% \\
22& 94~\% \\
23 & 100~\% \\
24 & 100~\% \\
\end{tabular}
\end{ruledtabular}
\end{table}

\begin{figure}
\centering
\includegraphics[scale=0.48]{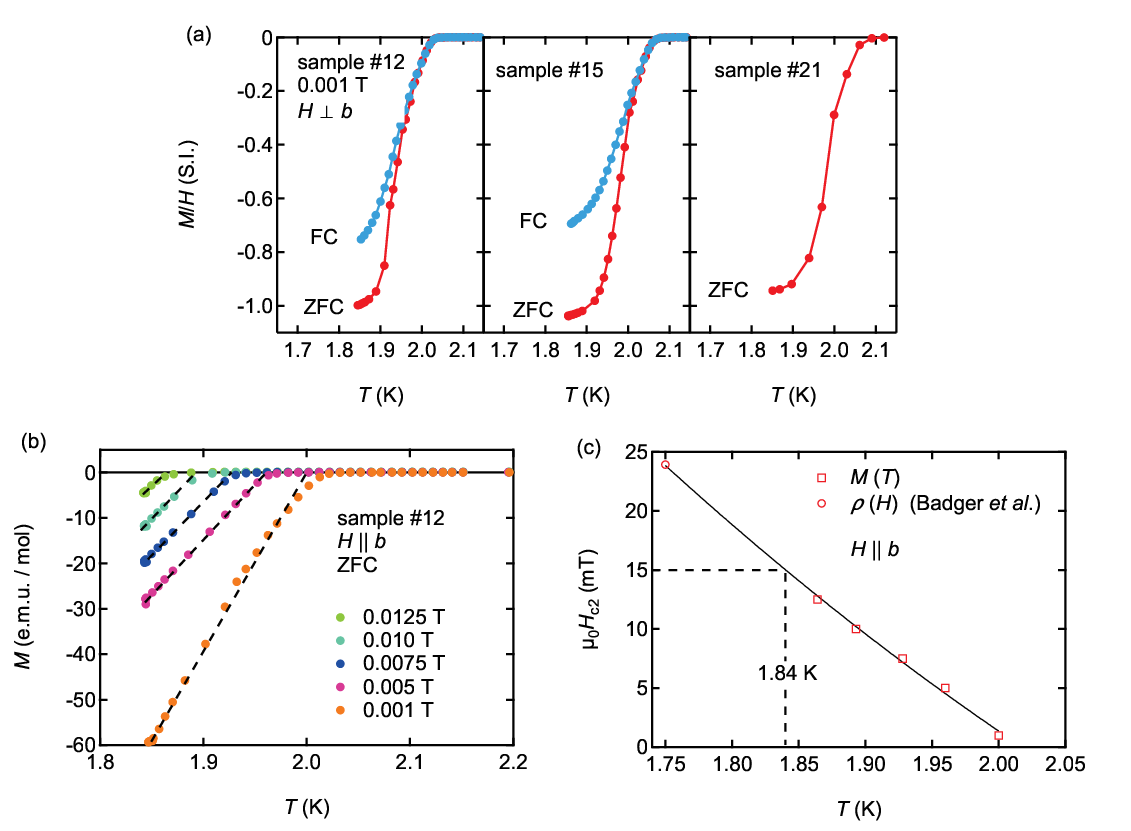}
\caption{(a) Temperature dependence of the bulk magnetic susceptibility of three of the LaNiGa$_2$ single crystals recorded under zero-field cooled (ZFC) and field-cooled (FC)
conditions. (b) Measurements of the bulk magnetic susceptibility of one of the single crystals for different values of magnetic field applied parallel to the $b$ axis
under ZFC conditions. (c) Temperature dependence of the upper critical magnetic field determined from the measurements in (b). The data point represented by 
an open circle is from resistivity measurements in Ref.~\cite{Badger:2022}. As shown, the value of $T_c$ at $H \! = \! 150$~Oe is estimated to be 1.84~K.}
\end{figure}

\noindent
{\bf Determination of $\lambda_{ac}$ from the Second Moment of the Internal Magnetic Field Distribution}\\

Provided $\lambda_{ac} \! \gg \! \xi_{ac}$, $\lambda_{ac}^{-4}$ is proportional to the vortex-lattice contribution to the second moment $\langle \Delta B^2 \rangle$ 
of the internal magnetic field distribution, $P(B)$, probed by $\mu$SR \cite{Brandt:88}. 
Following the method described in Ref.~\cite{Khasanov:23}, the TF-$\mu$SR asymmetry spectra were fit to the sum of four Gaussian-damped cosine functions
\begin{eqnarray} 
A(t) & = & \sum_{i = 1}^{3} a_{i} e^{-\sigma_{i}^2 t^2} \cos(2 \pi \nu_{i} t + \phi) \nonumber \\
      &     & +  a_{\rm bg}e^{-\sigma_{\rm bg}^2 t^2}\cos(2 \pi \nu_{\rm bg} t + \phi) \nonumber \, ,    
\label{eq:1}
\end{eqnarray}
where the first term denotes the contribution from the LaNiGa$_2$ sample and the second term denotes the background contribution from muons that did not stop in the sample.

The second moment of the internal magnetic field distribution of the LaNiGa$_2$ sample is obtained as follows
\begin{equation} 
\langle \Delta B^2 \rangle = \sum_{i = 1}^{3}\frac{a_i}{a_1 + a_2 + a_3} \left[ \left(\frac{\sigma_{i}}{\gamma_\mu}\right)^2 + (B_i - \langle B \rangle)^2   \right] \nonumber \, ,
\end{equation}
where $\gamma_\mu/2 \pi \! = \! 13.5539$~MHz/kG is the muon gyromagnetic ratio, $B_i \! = \! 2 \pi \nu_i/\gamma_\mu$ and $\langle B \rangle$ is the first moment given by
\begin{equation} 
\langle B \rangle = \sum_{i = 1}^{3}\frac{a_i B_i}{a_1 + a_2 + a_3} \nonumber \, .
\end{equation}
The vortex lattice (VL) contribution is obtained by subtracting the nuclear dipole contribution to $\langle \Delta B^2 \rangle$ as follows
\begin{equation}
\langle \Delta B^2 \rangle_{\rm VL} = \langle \Delta B^2 \rangle - \left(\frac{\sigma_{\rm n}}{\gamma_\mu}\right)^2 \nonumber \, .
\end{equation}
Assuming an hexagonal vortex lattice, $\lambda_{ac}$ is related to $\langle \Delta B^2 \rangle_{\rm VL}$ by the following equation
\begin{equation}
\lambda_{ac} [{\rm \AA}] = 10^4\sqrt{\frac{4.83(1 - b)[1 + 1.21(1 - \sqrt{1 - b})^3]}{\gamma_\mu \langle \Delta B^2 \rangle_{\rm VL}^{1/2}}} \nonumber \, ,
\end{equation}
where $b \! = \! B/B_{c2}$ is the reduced field for a field applied perpendicular to the $ac$ plane. 
Figure~S2 shows the temperature dependence of $\lambda_{ac}$ obtained in this way compared to the results obtained by the method described in the main article.

\begin{figure}
\centering
\includegraphics[scale=0.35]{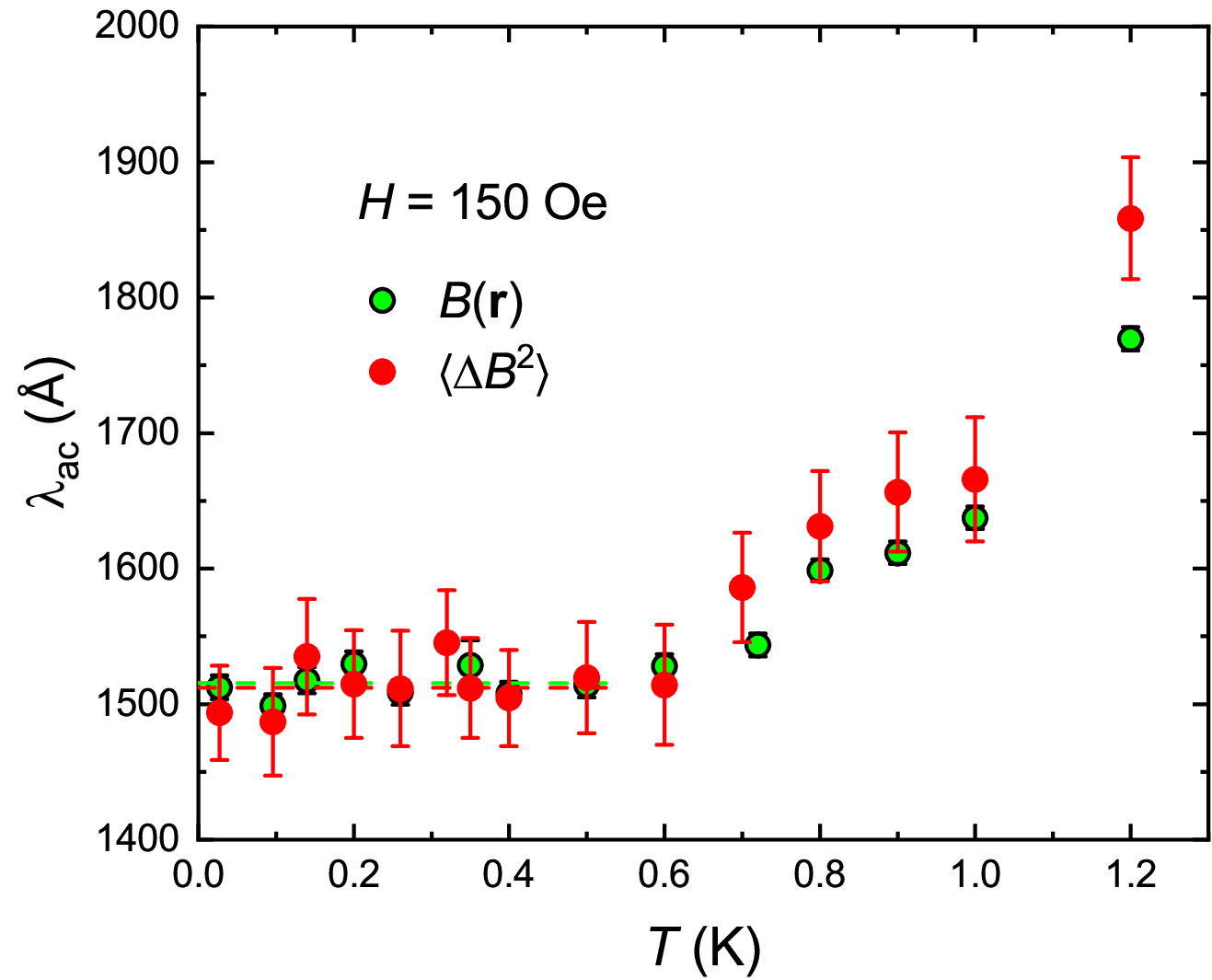}
\caption{Comparison of the temperature dependence of the magnetic penetration depth $\lambda_{ac}$ in LaNiGa$_2$ from fits of the TF-$\mu$SR asymmetry
spectra as described in the main article ({\it i.e.}, assuming an analytical model for the field profile $B({\bf r})$ generated by the vortex lattice) and calculated from the
second moment $\langle \Delta B^2 \rangle$ obtained from fits of the TF-$\mu$SR spectra to the sum of four Gaussian-damped cosine functions. The dashed horizontal lines
are fits of each data set below 0.5~K, which yield $\lambda_{ac}(0) \! = \! 1516 \! \pm \! 4$~\AA~ and $\lambda_{ac} \! = \! 1512 \! \pm \! 6$~\AA~ for the
two different analysis methods.}
\end{figure}

\bibliographystyle{apsrev}

\begin{thebibliography}{38}

\bibitem{Sarma:2015} S. Das  Sarma, M. Freedman, and C. Nayak, Majorana zero modes and topological quantum computation, npj Quantum Information {\bf 1}, 15001 (2015). 

\bibitem{Sato:2017} M. Sato, and Y. Ando, Topological superconductors: a review, Rep. Prog. Phys. {\bf 80}, 076501 (2017).

\bibitem{Badger:2022} J. R. Badger, Y. Quan, M. C. Staab, S. Sumita, A. Rossi, K. P. Devlin, K. Neubauer, D. S. Shulman, J. C. Fettinger, P. Klavins, S. M. Kauzlarich,
D. Aoki, I. M. Vishik, W. E. Pickett, and V. Taufour, Dirac lines and loop at the Fermi level in the time-reversal symmetry breaking superconductor 
LaNiGa$_2$, Commun. Phys. {\bf 5}, 22 (2022).

\bibitem{Hillier:2012} A. D. Hillier, J. Quintanilla, B. Mazidian, J. F. Annett, and R. Cywinski, Nonunitary triplet pairing in the centrosymmetric superconductor 
LaNiGa$_2$, Phys. Rev. Lett. {\bf 109}, 097001 (2012).

\bibitem{Weng:2016} Z. F. Weng, J. L. Zhang, M. Smidman, T. Shang, J. Quintalla, J. F. Annett, M. Nicklas, G. M. Pang, L. Jiao, W. B. Jiang, Y. Chen, F, Steglich, and H. Q. Yuan,
Two-gap superconductivity in LaNiGa$_2$ with nonunitary triplet pairing and even parity gap symmetry, Phys. Rev. Lett. {\bf 117}, 027001 (2016).

\bibitem{Csire:2018} G. Csire, B. {\'U}jfalussy, and J. F. Annett, Nonunitary triplet pairing in the noncentrosymmetric superconductor LaNiC$_2$, Eur. Phys. J. B {\bf 91}, 217 (2018).

\bibitem{Ghosh:2020} S. K. Ghosh, G. Csire, P. Whittlesea, J. F. Annett, M. Gradhand, B. {\'U}jfalussy, and J. Quintanilla, Quantitative theory of triplet pairing in the unconventional superconductor LaNiGa$_2$, Phys. Rev. B {\bf 101}, 100506(R) (2020).

\bibitem{Hillier:2009} A. D. Hillier, J. Quintanilla, and R. Cywinski, Evidence for time-reversal symmetry breaking in the noncentrosymmetric superconductor 
LaNiC$_2$, Phys. Rev. Lett. {\bf 102}, 117007 (2009).

\bibitem{Sundar:2021} S. Sundar, S. R. Dunsiger, S. Gheidi, K. S. Akella, A. M. C{\^o}t{\'e}, H. U. \"{O}zdemir, N. R. Lee-Hone, D. M. Broun, E. Mun, F. Honda, Y. J. Sato, 
T. Koizumi, R. Settai, Y. Hirose, I. Bonalde, and J. E. Sonier, Two-gap time reversal symmetry breaking superconductivity in noncentrosymmetric LaNiC$_2$, Phys. Rev. B {\bf 103}, 014511 (2021).

\bibitem{Lee:1996} W. H. Lee, H. K. Zeng, Y. D. Yao, and Y. Y. Chen, Superconductivity in the Ni based ternary carbide LaNiC$_2$, Physica C {\bf 266}, 138 (1996).

\bibitem{Pecharsky:1998} V. K. Pecharsky, L. L. Miller, and K. A. Gschneidner, Jr., Low-temperature behavior of two ternary lanthanide nickel carbides: Superconducting 
LaNiC$_2$ and magnetic CeNiC$_2$, Phys. Rev. B {\bf 58}, 497 (1998).

\bibitem{Iwamoto:1998} Y. Iwamoto, Y. Iwasaki, K. Ueda, and T. Kohara, Microscopic measurements in $^{139}$La-NQR of the ternary carbide superconductor LaNiC$_2$,
Phys. Lett. A {\bf 250}, 439 (1998).

\bibitem{Katano:2017} S. Katano, K. Shibata, K. Nakashima, and Y. Matsubara, Magnetic impurity effects on the superconductivity of noncentrosymmetric 
LaNiC$_2$: Ce substitution for La, Phys. Rev. B {\bf 95}, 144502 (2017).

\bibitem{Chen:2013} J. Chen, L. Jiao, J. L. Zhang, Y. Chen, L. Yang, M. Nicklas, F. Steglich, and H. Q. Yuan, Evidence for two-gap superconductivity in the
non-centrosymmetric compound LaNiC$_2$, New. J. Phys. {\bf 15}, 053005 (2013).

\bibitem{Bonalde:2011} I. Bonalde, R. L. Ribeiro, K. J. Syu, H. H. Sung, and W. H. Lee, 
Nodal gap structure in the noncentrosymmetric superconductor LaNiC$_2$ from magnetic-penetration-depth measurements, New. J. Phys. {\bf 13}, 123022 (2011).     

\bibitem{Landaeta:2017} J. F. Landaeta, D. Subero, P. Machado, F. Honda, and I. Bonalde, Unconventional superconductivity and an ambient-pressure magnetic quantum critical point in single-crystal LaNiC$_2$, Phys. Rev. B {\bf 96}, 174515 (2017).

\bibitem{SM} See Supplemental Material for bulk magnetic susceptibility measurements of the superconducting shielding fraction, $T_c$ and the upper critical magnetic field $H_{c2}(T)$,
and for a determination of $\lambda_{ac}$ from fits of the TF-$\mu$SR spectra for $H \! = \! 150$~Oe below $T_c$ to the sum of four Gaussian-damped cosine functions.
The Supplemental Material also contains Ref.~\cite{Khasanov:23}.

\bibitem{Khasanov:23} R. Khasanov, A. Ramires, V. Grinenko, I. Shipulin, N. Kikugawa, D. A. Sokolov, J. A. Krieger, T. J. Hicken, Y. Maeno, H. Luetkens, and Z. Guguchia,
In-Plane Magnetic Penetration Depth in Sr$_2$RuO$_4$: Muon-spin rotation and relaxation study, Phys. Rev. Lett. {\bf 131}, 236001 (2023).  

\bibitem{Sonier:2000} J. E. Sonier, J. H. Brewer and R. F. Kiefl, $\mu$SR studies of the vortex state in type-II superconductors, Rev. Mod. Phys. {\bf 72}, 769 (2000).

\bibitem{Sonier:2004} J. E. Sonier, F. D. Callaghan, R. I. Miller, E. Boaknin, L. Taillefer, R. F. Kiefl, J. H. Brewer, K. F. Poon, and J. D. Brewer, Shrinking magnetic vortices
in V$_3$Si due to delocalized quasiparticle core states: Confirmation of the microscopic theory for interacting vortices, Phys. Rev. Lett. {\bf 93}, 017002 (2004).

\bibitem{Sonier:2007} J. E. Sonier, Muon spin rotation studies of electronic excitations and magnetism in the vortex cores of superconductors, Rep. Prog. Phys. {\bf 70}, 1717 (2007).

\bibitem{Takanaka:1971} K. Takanaka, Flux-line lattices of pure type II superconductors with anisotropic Fermi surface, Prog. Theor. Phys. {\bf 46}, 1301-1306 (1971).

\bibitem{BCS:1957} J. Bardeen, L. N. Cooper, and J. R. Schrieffer, Microscopic theory of superconductivity, Phys. Rev. {\bf 106}, 162 (1957).

\bibitem{Brandt:88} E. H. Brandt, Flux distribution and penetration depth measured by muon spin rotation in high-$T_c$ superconductors, Phys. Rev. B {\bf 37}, R2349 (1988).

\bibitem{Kogan:2009} V. G. Kogan, C. Martin, and R. Prozorov, Superfluid density and specific heat within a self-consistent scheme for a two-band superconductor. 
Phys. Rev. B {\bf 80}, 014507 (2009).

\bibitem{Kadono:2004} R. Kadono, Field-induced quasiparticle excitations in novel type II superconductors, J. Phys.: Condens. Matter {\bf 16}, S4421 (2004).

\bibitem{Ohishi:2002} K. Ohishi, K. Kahuta, J. Akimitsu, W. Higemoto, R. Kadono, J.E. Sonier, A.N. Price, R.I. Miller, R.F. Kiefl, M. Nohara, H. Suzuki, and H. Takagi,
Nonlocal effects and shrinkage of the vortex core radius in YNi$_2$B$_2$C probed by muon spin rotation, Phys. Rev. B {\bf 65}, 140505(R) (2002).

\bibitem{Price:2002} A.N. Price, R.I. Miller, R.F. Kiefl, J.A. Chakhalian, S.R. Dunsiger, G.D. Morris, J.E. Sonier, and P.C. Canfield,
Anomalous vortex state of superconducting LuNi$_2$B$_2$C, Phys. Rev. B {\bf 65}, 214520 (2002).

\bibitem{Callaghan:2005} F. D. Callaghan, M. Laulajainen, C. V. Kaiser, and J. E. Sonier, Field dependence of the vortex core size in a multiband superconductor,
Phys. Rev. Lett. {\bf 95}, 197001 (2005). 

\bibitem{Williams:2009} T. J. Williams, A. A. Aczel, E. Baggio-Saitovitch, S. L. Bud'ko, P. C. Canfield, J. P. Carlo, T. Goko, J. Munevar,
N. Ni, Y. J. Uemura, W. Yu, and G. M. Luke, Muon spin rotation measurement of the magnetic field penetration depth in Ba(Fe$_{0.926}$Co$_{0.074}$)2As$_2$:
Evidence for multiple superconducting gaps, Phys. Rev. B {\bf 80}, 094501 (2009).

\bibitem{Weyeneth:2010} S. Weyeneth, M. Bendele, R. Puzniak, F. Mur\'{a}nyi, A. Bussmann-Holder, N. D. Zhigadlo, S. Katrych, Z. Bukowski, J. Karpinski, A. Shengelaya,
Field-dependent superfluid density in the optimally doped SmFeAsO$_{1-x}$F$_y$ superconductor, Europhys. Lett. {\bf 91}, 47005 (2010).  

\bibitem{Khasanov:2008} Rustem Khasanov, Hubertus Luetkens, Alex Amato, Hans-Henning Klauss, Zhi-An Ren, Jie Yang, Wei Lu, and Zhong-Xian Zhao, Muon spin rotation studies of 
SmFeAsO$_{0.85}$ and NdFeAsO$_{0.85}$ superconductors, Phys. Rev. B {\bf 78}, 092506 (2008).

\bibitem{Khasanov:2007} R. Khasanov, A. Shengelaya, A. Maisuradze, F. La Mattina, A. Bussmann-Holder, H. Keller, and K. A. M\"{u}ller,
Experimental evidence for two gaps in the high-temperature La$_{1.83}$Sr$_{0.17}$CuO$_4$ superconductor, Phys. Rev. Lett. {\bf 98}, 057007 (2007).

\bibitem{Serventi:2004} S. Serventi, G. Allodi, R. De Renzi, G. Guidi, and L. Roman\`{o}, P. Manfrinetti, A. Palenzona, Ch. Niedermayer, A. Amato, and Ch. Baines,
Effect of two gaps on the flux-lattice internal field distribution: Evidence of two length scales in Mg$_{1-x}$Al$_x$B$_2$ from $\mu$SR, Phys. Rev. Lett. {\bf 93}, 217003 (2004).

\bibitem{Luetkens:2008} H. Luetkens, H.-H. Klauss, R. Khasanov, A. Amato, R. Klingeler, I. Hellmann, N. Leps, A. Kondrat, C. Hess, A. K\"{o}hler, G. Behr, J. Werner, and B. B\"{u}chner,
Field and temperature dependence of the superfluid density in LaFeAsO$_{1-x}$F$_x$ superconductors: A muon spin relaxation study, Phys. Rev. Lett. {\bf 101}, 097009 (2008).

\bibitem{Cho:2022} Kyuil Cho, M. Ko\'{n}czykowski, S. Ghimire, M. A. Tanatar, Lin-Lin Wang, V. G. Kogan, and R. Prozorov, Multi-band s++ superconductivity in V$_3$Si determined
from the response to a controlled disorder, Phys. Rev. B {\bf 105}, 024506 (2022).

\bibitem{Ding:2023} S. Ding, D. Zhao, T. Jiang, H. Wang, D. Feng, and T. Zhang, Surface structure and multigap superconductivity of V3Si (111) revealed by scanning tunneling microscopy, Quantum Frontiers {\bf 2}, 3 (2023).

\bibitem{Sonier:1997} J. E. Sonier, R. F. Kiefl, J. H. Brewer, J. Chakhalian, S. R. Dunsiger, W. A. MacFarlane, R. I. Miller, A. Wong, G. M. Luke, and J. W. Brill,
Muon-spin rotation measurements of the magnetic field dependence of the vortex-core radius and magnetic penetration depth in NbSe$_2$, Phys. Rev. Lett. {\bf 79}, 1742 (1997).

\bibitem{Ichioka:1999a} M. Ichioka, A. Hasegawa, and K. Machida, Vortex lattice effects on low-energy excitations in $d$-wave and $s$-wave superconductors,
Phys. Rev. B {\bf 59}, 184 (1999).  

\bibitem{Ichioka:1999b} M. Ichioka, A. Hasegawa, and K. Machida, Field dependence of the vortex structure in $d$-wave and $s$-wave superconductors,
Phys. Rev. B {\bf 59}, 8902 (1999).

\bibitem{Boaknin:2003} E. Boaknin, M. A. Tanatar, J. Paglione, D. Hawthorn, F. Ronning, R. W. Hill, M. Sutherland, L. Taillefer, J. Sonier, S. M. Hayden, and J. W. Brill,
Heat conduction in the vortex state of NbSe$_2$: Evidence for multiband superconductivity, Phys. Rev. Lett. {\bf 90}, 117003 (2003).

\bibitem{Kogan:2005} V. G. Kogan and N. V. Zhelezina, Field dependence of the vortex core size, Phys. Rev. B {\bf 71}, 134505 (2005).

\bibitem{Silaev:2011} M. Silaev and E. Babaev, Microscopic theory of type-1.5 superconductivity in multiband systems, Phys. Rev. B {\bf 84}, 094515 (2011).

\bibitem{Ichioka:2017} M. Ichioka, V. G. Kogan, and J. Schmalian, Locking of length scales in two-band superconductors, Phys. Rev. B {\bf 95}, 064512 (2017).

\bibitem{Vargunin:2019} A. Vargunin and M. A. Silaev, Field dependence of the vortex-core size in dirty two-band superconductors, Phys. Rev. B {\bf 100}, 014516 (2019).    

\end{thebibliography}

\begin{thebibliography}{38}
\bibitem{Badger:2022} J. R. Badger, Y. Quan, M. C. Staab, S. Sumita, A. Rossi, K. P. Devlin, K. Neubauer, D. S. Shulman, J. C. Fettinger, P. Klavins, S. M. Kauzlarich,
D. Aoki, I. M. Vishik, W. E. Pickett, and V. Taufour, Dirac lines and loop at the Fermi level in the time-reversal symmetry breaking superconductor 
LaNiGa$_2$, Commun. Phys. {\bf 5}, 22 (2022).\\

\bibitem{Brandt:88} E. H. Brandt, Flux distribution and penetration depth measured by muon spin rotation in high-$T_c$ superconductors, Phys. Rev. B {\bf 37}, R2349 (1988).\\

\bibitem{Khasanov:23} R. Khasanov, A. Ramires, V. Grinenko, I. Shipulin, N. Kikugawa, D. A. Sokolov, J. A. Krieger, T. J. Hicken, Y. Maeno, H. Luetkens, and Z. Guguchia,
In-Plane Magnetic Penetration Depth in Sr$_2$RuO$_4$: Muon-spin rotation and relaxation study, Phys. Rev. Lett. {\bf 131}, 236001 (2023).
\end{thebibliography}

\end{document}